\documentclass[12pt]{article}
\usepackage{amsmath, amssymb,graphicx}

\newwrite\ffile\global\newcount\figno \global\figno=1

\def\writedef#1{}

\input epsf
\def\figin{\epsfcheck\figin}\def\figins{\epsfcheck\figins}
\def\epsfcheck{\ifx\epsfbox\UnDeFiNeD
\message{(NO epsf.tex, FIGURES WILL BE IGNORED)}
\gdef\figin##1{\vskip2in}\gdef\figins##1{\hskip.5in}
\else\message{(FIGURES WILL BE INCLUDED)}%
\gdef\figin##1{##1}\gdef\figins##1{##1}\fi}

\def\figinsert{}
\def\ifig#1#2#3{\xdef#1{fig.~\the\figno}
\writedef{#1\leftbracket fig.\noexpand~\the\figno}%
\figinsert\figin{\centerline{#3}}\medskip\centerline{\vbox{\baselineskip12pt
\advance\hsize by -1truein\center\footnotesize{  Fig.~\the\figno.} #2}}
\bigskip\endinsert\global\advance\figno by1}
\def\endinsert{}

\usepackage{amssymb}
\usepackage{graphics}
\begin{document}
\baselineskip 18pt
\newcommand{\Tr}{\mbox{Tr\,}}
\newcommand{\beq}{\begin{equation}}
\newcommand{\eeq}{\end{equation}}
\newcommand{\bea}{\begin{eqnarray}}
\newcommand{\eea}[1]{\label{#1}\end{eqnarray}}
\renewcommand{\Re}{\mbox{Re}\,}
\renewcommand{\Im}{\mbox{Im}\,}
\newcommand{\yms}{${YM^*\,}$}

\def\N{{\cal N}}


\thispagestyle{empty}
\renewcommand{\thefootnote}{\fnsymbol{footnote}}

{\hfill \parbox{4cm}{
        SHEP-09-27 \\
}}

\bigskip

\begin{center} \noindent \Large \bf
Holographic Integral Equations and Walking Technicolour
\end{center}

\bigskip\bigskip\bigskip

\centerline{ \normalsize \bf Raul Alvares, Nick Evans, Astrid
Gebauer \& George James Weatherill \footnote[1]{\noindent \tt
rpa1e08@soton.ac.uk, evans@soton.ac.uk, ag806@soton.ac.uk,
 gjw@soton.ac.uk} }

\bigskip
\bigskip\bigskip

\centerline{ \it School of Physics and Astronomy} \centerline{ \it
Southampton University} \centerline{\it  Southampton, SO17 1BJ }
\centerline{ \it United Kingdom}
\bigskip

\bigskip\bigskip

\renewcommand{\thefootnote}{\arabic{footnote}}

\centerline{\bf \small Abstract} We study chiral symmetry breaking
in the holographic D3-D7 system in a simple model with an
arbitrary running coupling. We derive equations for the D7
embedding and show there is a light pion. In particular we present
simple integral equations, involving just the running coupling and
the quark self energy, for the quark condensate and the pion decay
constant. We compare these to the Pagels-Stokar or constituent
quark model equivalent. We discuss the implications for walking
Technicolour theories. We also perform a similar analysis in the
four dimensional field theory whose dual is the non-supersymmetric
D3-D5 system and propose that it represents a walking theory in
which the quark condensate has dimension $2 + \sqrt{3}$.
\newpage

\section{Introduction}

The D3-D7 system\cite{Polchinski,Bertolini:2001qa,Karch} in
AdS-like spaces has allowed the study, through gauge/gravity
duality or holography\cite{Malda,Witten:1998qj,Gubser:1998bc}, of
many aspects of strongly interacting gauge theories with
quarks\cite{Erdmenger:2007cm}. The system has been used to study
quark confinement\cite{Maldacena:1998im,Rey:1998ik},
mesons\cite{Mateos}, transport properties at finite
temperature\cite{Son:2007vk,Gubser:2009md,Mateos:2007vn}, and
chiral symmetry breaking in the presence of a running
coupling\cite{Babington,Ghoroku:2004sp,Kruczenski:2003uq} or a
magnetic field\cite{Filev:2007gb}.

Here we wish to present a very simple model of chiral symmetry
breaking and the associated Goldstone boson (essentially pion) in
this system. The simple model consists of embedding the D7s in
pure $AdS_5 \times S^5$ but with an arbitrary dilaton profile to
represent the running coupling of the dual gauge theory. This
basic model, although the metric is not back reacted to the
dilaton's presence, provides a simple encapsulation of the chiral
symmetry breaking mechanism in the D3-D7 system. In particular it
will allow us to elucidate in the holographic equations of motion
why there is a Goldstone boson present for the symmetry breaking.
Further it will allow us to write integral equations for the
parameters of the low energy chiral Lagrangian involving just the
form of the running coupling and the quark self energy function
(the D7 brane embedding function). These equations are very
similar in spirit to the Pagels-Stokar formula\cite{Pagels:1979hd}
for the pion decay constant, $f_\pi$, and constituent quark
model\cite{Holdom:1990iq} estimates of the chiral condensate and
so forth.

The formulae we will present for these low energy parameters allow
one to develop intuition about how the low energy theory depends
on the underlying gauge dynamics. We explore this and as a
particular example look at
walking\cite{Holdom:1981rm,Appelquist:1986an}
Technicolour\cite{Weinberg:1975gm,Susskind:1978ms} theories to see
if the holographic model matches the folk lore from constituent
quark models. Our results support the expectation that a walking
regime will enhance the quark condensate relative to the pion
decay constant.

In our final section we will perform a similar study for the
non-supersymmetric D3/D5 system with a four dimensional overlap.
We interpret this system as a walking gauge theory where the quark
condensate has a dimension of $2+\sqrt{3}$ in the far UV. This
theory is not of any obvious phenomenological use but the walking
paradigm does seem to explain the physics of the system.

\section{A simple D3-D7 chiral symmetry breaking model}

We will consider a gauge theory with a holographic dual described
by the Einstein frame geometry $AdS_5 \times S^5$ \beq ds^2 = {1
\over g_{uv}} \left[{r^2 \over R^2} dx_{4}^2 + {R^2 \over r^2}
\left( d\rho^2 + \rho^2 d\Omega_3^2 + dw_5^2 + dw_6^2
\right)\right]\eeq where we have split the coordinates into the
$x_{3+1}$ of the gauge theory, the $\rho$ and $\Omega_3$ which
will be on the D7 brane world-volume and two directions transverse
to the D7, $w_5,w_6$. The radial coordinate, $r^2 = \rho^2 + w_5^2
+ w_6^2$, corresponds to the energy scale of the gauge theory. The
radius of curvature is given by $R^4 = 4 \pi g_{uv}^2 N
\alpha^{'2}$ with $N$ the number of colours. $g_{uv}^2$ is the $r
\rightarrow \infty$ value of the dilaton. In addition we will
allow an arbitrary running as $r \rightarrow 0$ to represent the
gauge theory coupling \beq e^{\phi} ~~ = ~~ g^2_{YM}(r^2)~~~  = ~~
g^2_{uv} ~\beta(\rho^2 + w_5^2 + w_6^2)  \eeq where the function
$\beta \rightarrow 1$ as $r \rightarrow \infty$. The $r
\rightarrow \infty$ limit of this theory is dual to the ${\cal
N}=4$ super Yang-Mills theory and $g^2_{uv}$ is the constant large
$r$ asymptotic value of the gauge coupling.

We will introduce a single D7 brane probe\cite{Karch} into the
geometry to include quarks - by treating the D7 as a probe we are
working in a quenched approximation although we can reintroduce
some aspects of quark loops through the running coupling's form if
we wish. Although this system only has a U(1) axial symmetry on
the quarks corresponding to rotations in the $w_5-w_6$ plane
(formally this symmetry is an R-symmetry of the model but it is
broken by a quark mass or condensate) we believe it is a good
setting for studying the dynamics of the quark condensation. That
process is driven by the strong dynamics rather than the global
symmetries so the absence of a non-abelian axial symmetry should
not be important\footnote{The Sakai Sugimoto
model\cite{Sakai:2004cn} is an example of a gravity dual with a
non-abelian chiral symmetry but it is fundamentally five
dimensional and a clear prescription for including a quark mass is
lacking - the result is that we would not know how to do this
analysis in that model since we can not identify the quark self
energy nor the quark condensate.}.

We must find the D7 embedding function eg $w_5(\rho), w_6=0$. The
Dirac Born Infeld action in Einstein frame is given by \beq
\begin{array}{ccl}
S_{D7} & = & -T_7 \int d^8\xi e^\phi  \sqrt{- \det P[G]_{ab}}\\ &&\\
&=&  -\overline{T_7} \int d^4x~ d \rho ~ \rho^3 \beta \sqrt{1 +
(\partial_\rho w_5)^2} \end{array} \eeq where $T_7 = 1/(2 \pi)^7
\alpha^{'4}$ and $\overline{T_7} = 2 \pi^2 T_7/ g^2_{uv}$ when we
have integrated over the 3-sphere on the D7. The equation of
motion for the embedding function is therefore \beq \label{embed}
\partial_\rho \left[ {\beta \rho^3
\partial_\rho w_5 \over \sqrt{1+ (\partial_\rho w_5)^2}}\right] - 2 w_5 \rho^3
\sqrt{1+ (\partial_\rho w_5)^2} {\partial \beta \over \partial
r^2} = 0 \eeq The UV asymptotic of this equation, provided the
dilaton returns to a constant so the UV dual is the ${\cal N}=4$
super Yang-Mills theory, has solutions of the form \beq
\label{asy}w_5 = m + {c \over \rho^2} +... \eeq where we can
interpret $m$ as the quark mass ($m_q = m/2 \pi \alpha'$) and $c$
is proportional to the quark condensate as we'll see below.

The embedding equation (\ref{embed}) clearly has regular solutions
$w_5=m$ when $g^2_{YM}$ is independent of $r$ - the flat
embeddings of the ${\cal N}=2$ Karch-Katz theory\cite{Karch}.
Equally clearly if $\partial \beta / \partial r^2$ is none trivial
in $w_5$ then the second term in (\ref{embed}) will not vanish for
a flat embedding. We conclude that for any non-trivial gauge
coupling the asymptotic solutions must contain the parameter $c$,
a quark condensate. Whether $c \rightarrow 0$ or not as $m
\rightarrow 0$ depends on the precise form of the running coupling
chosen (note that $w_5=0$ is always a solution of (\ref{embed})).
However, if the coupling grows towards $r=0$ as one would expect
in a confining theory then there is clearly a growing penalty in
the action for the D7 to approach the origin and we expect $c$ to
be non-zero.

As an example one can consider a gauge coupling running with a
step of the form \beq \label{coupling}\beta =  a+1 - a \tanh\left[
\Gamma (r - \lambda) \right] \eeq  This form introduces conformal
symmetry breaking at the scale $\Lambda = \lambda/2 \pi \alpha'$
which triggers chiral symmetry breaking. The parameter $a$
determines the increase in the coupling across the step but the
solutions have only a small dependence on the value chosen because
the area of increasing coupling is avoided by the D7 brane. An
extreme choice of the profile is to let the coupling actually
diverge at the barrier to represent the one loop blow up in the
running of the QCD coupling - the solutions show the same
behaviour as for a finite step provided the transition is not
infinitely sharp. The parameter $\Gamma$ spreads the increase in
the coupling over a region in $r$ of order $\Gamma^{-1}$ in size -
the effect of widening the step is to enhance the tail of the self
energy function for the quark. We show the symmetry breaking
embeddings in Figure 1.

\begin{center}
\center{\includegraphics[width=80mm]{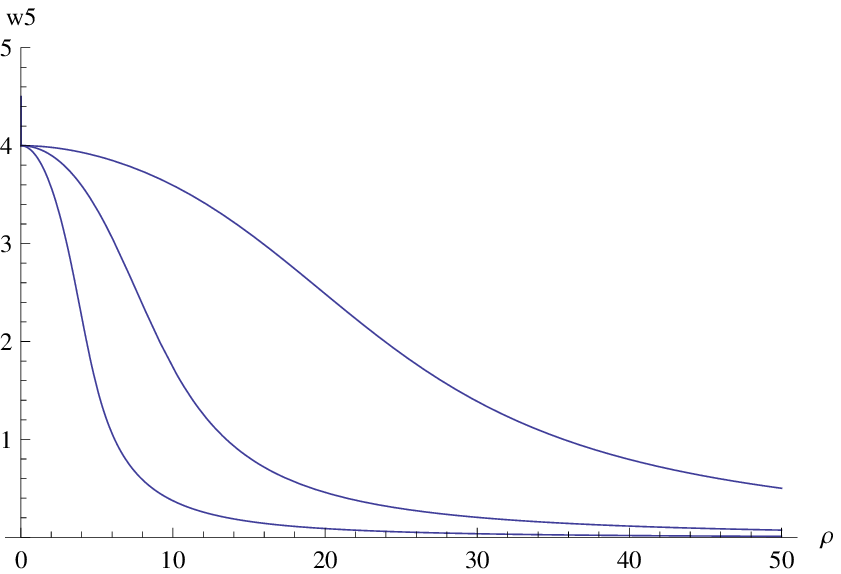}}
\end{center}

{\small{Figure 1: The D7 brane embeddings/quark self energy plots
for the coupling ansatz in (\ref{coupling}) - in each case the
parameter $a=3$ and from left to right: $\lambda = 3.19,
\Gamma=1$; $\lambda = 4.55, \Gamma=0.3$; $\lambda = 10.4,
\Gamma=0.1$.}} \vspace{1cm}

We will interpret the D7 embedding function as the dynamical self
energy of the quark, similar to that emerging from a gap equation.
The separation of the D7 from the $\rho$ axis is the mass at some
particular energy scale given by $\rho$ - in the ${\cal N}=2$
theory where the embedding is flat the mass is not renormalized,
whilst with the running coupling an IR mass forms - we have picked
parameters in Figure 1 that generate the same dynamical quark mass
at $\rho=0$. We call the embedding function $\Sigma_0$ below.

\subsection{Goldstone Mode}

The embedding above lies at $w_6=0$ but there is clearly a set of
equivalent solutions given by rotating that solution in the
$w_5-w_6$ plane. That degeneracy of the solutions is the vacuum
manifold. We therefore expect a Goldstone mode associated with a
fluctuation of the vacuum in the angular direction. For small
fluctuations about the embedding above we may look at fluctuations
in $w_6$. The quadratic order expanded action for such a small
fluctuation is
\begin{eqnarray} \label{w6act}  S_7&=&-\overline{T_7}\int \mathrm{d}\rho  dx^4
\rho^3\beta \sqrt{1+(\partial_\rho \Sigma_0)^2}\left( 1+
{\partial_{r^2}\beta \over \beta} w_6^2 +
    \frac{1}{2}\frac{(\partial_\rho w_6)^2} {1+(\partial_\rho
\Sigma_0)^2} + {1 \over 2} {R^4 \over r^4} (\partial_x w_6)^2 +...
\right)
\end{eqnarray} note $r$, $\beta$ and $\partial_{r^2} \beta$ are
evaluated on the solution $\Sigma_0$ here and henceforth.

As usual we will seek solutions  of the form
$w_6(\rho,x)=f_n(\rho)e^{ik.x}, ~~ k^2=-M_n^2$. Here $n$ takes
integer values - the solutions are associated with the Goldstone
boson and its tower of radially excited states. The $f_n$ satisfy
the equation
\begin{equation}
\partial_\rho(\frac{\beta \rho^3 \partial_\rho f_n}{\sqrt{1+(\partial_\rho \Sigma_0)^2}})
-2\rho^3\sqrt{1+(\partial_\rho \Sigma_0)^2} ( \partial_{r^2} \beta
) f_n+\frac{1}{r^4}\rho^3\beta \sqrt{1+(\partial_\rho
\Sigma_0)^2}R^4M_n^2f_n=0 \label{eq:equation2}
\end{equation}
The presence of a Goldstone boson is now immediately apparent -
there is a solution with $M_n^2=0$ and $f_0= \Sigma_0$. With these
substitutions the equation is exactly the embedding equation
(\ref{embed}), a result of the symmetry between between $w_5$ and
$w_6$. This is the pion like bound state of this theory - although
there is only a broken U(1) axial symmetry, the absence of anomaly
effects at large N make it closer in nature to the pions than the
$\eta'$ of QCD.

Naively the argument just given makes it appear there is a
massless Goldstone for any $w_5$ solution including those where
there is an explicit quark mass in the asymptotic fall off in
(\ref{asy}). This is not the case though because to interpret the
solution as a Goldstone requires $f_0$ to fall off at large $\rho$
as $1/\rho^2$ - it must be a fluctuation in the condensate not the
explicit mass. The naive massless solution is related to the fact
that the theory has a spurious symmetry where $\bar{\psi}_L \psi_R
\rightarrow e^{i \alpha} \bar{\psi}_L \psi_R$ and simultaneously
$m \rightarrow e^{-i \alpha} m$. This spurious symmetry must be
present in the string construction.

\subsection{The low energy Chiral Lagrangian}

The Goldstone field's low energy Lagrangian must take the form of
a chiral Lagrangian, non-linear realization of the broken
symmetry\cite{Coleman:1969sm}. We can substitute the form $w_6 =
f_0(\rho) \Pi(x)= \Sigma_0 \Pi(x)$ into (\ref{w6act}) and
integrate over $\rho$ to obtain this Lagrangian
\begin{eqnarray} \nonumber {\cal L}&=&-\overline{T_7}\int \mathrm{d}\rho
\rho^3 \beta \sqrt{1+(\partial_\rho \Sigma_0)^2} \left(1  +
\frac{1}{2}\frac{R^4}{r^4}\Sigma_0^2(\partial_x\Pi)^2 \right. \\
&& \left. \hspace{2cm}+ \frac{1}{4}\frac{R^4}{r^4}\left[{2\over
\beta} {d \beta \over d r^2} \Sigma_0^4 +
{\Sigma_0^2(\partial_\rho \Sigma_0)^2
 \over 1 + (\partial_\rho \Sigma_0)^2} \right](\partial_x\Pi)^2  \Pi^2 + ... \right)\label{eq:action3}
\end{eqnarray} where we've used the equation of motion
(\ref{eq:equation2}) to eliminate the second and third terms in
(\ref{w6act}) in the massless limit. We have also included the
$\Pi^2 (\partial_x \Pi)^2$ term from the fourth order expansion
from which we will determine $f_\pi$.

This should be compared to the standard chiral Lagrangian form
where $U =  {\rm exp}(\sqrt{2} i \pi/f_\pi)$
\begin{eqnarray}\label{chiral} {\cal L}& =
& V_0 + {f_\pi^2 \over 4}
\partial_x U^\dagger \partial^x U + {\cal O}(p^4)\nonumber  \\ &&\\
& = & V_0 + {1 \over 2} ( \partial_x \pi)^2 + {1 \over f_\pi^2}
(\partial_x \pi)^2 \pi^2 + {\cal O}(\pi^6) + {\cal O}(p^4)
\nonumber\end{eqnarray} where $V_0$ is the vacuum energy and
$f_\pi$ is the pion decay constant.

We must rescale $\Pi$ in (\ref{eq:action3}) to the canonical
normalization in (\ref{chiral}) and then we can read off an
integral expression for the pion decay constant. To ensure all
factors of $\alpha'$ are absent from physical answers, as they
must be, we must express our answer as the ratio of two physical
scales. Here we will use the scale $\Lambda$ in the gauge coupling
running (\ref{coupling}) that encodes the scale of the chiral
symmetry breaking as our reference - we have\beq {f_\pi^2 \over
\Lambda^2} = {-N \over  \pi^2 \lambda^2} \left. { \left[ \int d
\rho \rho^3 \beta \sqrt{  1 + (\partial_\rho \Sigma_0)^2}
{\Sigma_0^2 \over (\rho^2 + \Sigma_0^2)^2} \right]^2 \over \left[
\int d \rho \rho^3 \beta \sqrt{1+(\partial_\rho \Sigma_0)^2}
\frac{1}{4(\rho^2 + \Sigma_0^2)^2}\left({2\over \beta} {d \beta
\over d r^2} \Sigma_0^4 + {\Sigma_0^2(\partial_\rho \Sigma_0)^2
 \over 1 + (\partial_\rho \Sigma_0)^2}
\right) \right]} \right|_{r^2=\rho^2+\Sigma_0^2}
 \eeq
Note that $\partial_{r^2} \beta$ is typically negative for the
embeddings we have explored above so that $f_\pi^2$ is positive.
Employing the embedding equation (\ref{embed}) the denominator may
be simplified leaving \beq \label{fpi}{f_\pi^2 \over \Lambda^2} =
{- 4 N \over \pi^2 \lambda^2}  { \left[ \int d \rho \rho^3 \beta
\sqrt{  1 + (\partial_\rho \Sigma_0)^2} {\Sigma_0^2 \over (\rho^2
+ \Sigma_0^2)^2} \right]^2 \over \left[ \int d \rho
\frac{\Sigma_0^2}{(\rho^2 + \Sigma_0^2)^2}
\partial_\rho \left(
 { \beta \rho^3 \Sigma_0 (\partial_\rho
\Sigma_0) \over \sqrt{  1 + (\partial_\rho \Sigma_0)^2}}\right)
 \right]}  \eeq

 We can also extract an integral equation for the quark condensate
 (evaluated in the UV where there is no running) from our
 analysis. We use the fact that the expectation value of $\bar{q}_L q_R$
 is given by ${1 \over Z} {\partial Z \over \partial m_q}|_{m_q \rightarrow
 0}$. For an infinitessimal value of $m$ in the boundary embedding
 (\ref{asy}) we expect the full embedding, to leading order, to simply take
 the form $w_5 = 2 \pi \alpha' m_q + \Sigma_0$. We insert this form into the vacuum energy
 and expand to leading order in $m_q$ - the coefficient is just the
 quark condensate
\beq \label{cond}{\langle \bar{q}_L q_R \rangle \over \Lambda^3} =
{ -N \over 4 \pi \lambda^3 g^2_{uv} N} \left.
 \int d\rho ~\rho^3 \Sigma_0 \sqrt{ 1 + (
\partial_\rho \Sigma_0)^2} \partial_{r^2} \beta \right|_{r^2=\rho^2+\Sigma_0^2} \eeq
One may use the embedding equation (\ref{embed}) to turn this into
a surface term that is then, given $\beta$ becomes unity
asymptotically, proportional to $\rho^3 \partial_\rho
\Sigma_0|_{\rho \rightarrow \infty}$ which is just proportional to
the constant $c$ in (\ref{asy}) confirming the interpretation of
$c$ as the condensate. The integral form of the equation though
allows intuition for the value of the condensate from the shape of
the embedding as we will see. Note that if the 'tHooft coupling
$g^2_{uv} N$ is kept fixed both $f_\pi$ and the condensate grow as
$N$ as expected.

The integral equations (\ref{fpi}) and (\ref{cond}) that link low
energy parameters to the underlying UV physics are our main
results. They are very reminiscent of constituent quark
model\cite{Holdom:1990iq} results which input the quark self
energy, $\Sigma(q)$, (for example determined from a gap
equation\cite{Appelquist:1986an}) to determine the same
quantities. In particular those models give for the condensate
\beq \label{cond2} \langle \bar{q} q \rangle = {N \over 2} \int
q^3 dq {\Sigma \over q^2 + \Sigma^2} \eeq  and the Pagels Stokar
formula\cite{Pagels:1979hd} for the pion decay constant \beq
\label{fpi2} f_\pi^2 = {N \over 8 \pi^2} \int q^3 dq {\Sigma^2 -
{1 \over 2} q^2 \Sigma \Sigma' \over (q^2 + \Sigma^2)^2} \eeq
where a prime indicates a derivatives with respect to $q^2$.
Although our formulae are more complex and include the underlying
gauge coupling's running there are nevertheless a number of common
features. We will compare them for the case of walking
Technicolour below.

It must be stressed that we have derived our expressions
(\ref{fpi}) and (\ref{cond}) in a toy holographic model of chiral
symmetry breaking. Of course one can not just impose any random
running of the gauge coupling and assume one is in a real gauge
theory. We have also not included any back reaction of the space's
metric to the presence of a non-trivial dilaton. The analysis is
very similar in spirit to the chiral quark model assumption of an
arbitrary choice of $\Sigma(q^2)$. Despite these flaws, we hope
the simplicity of the expressions allows one to analytically
understand the typical response of the holographic descriptions to
different types of running coupling.

\section{Walking Technicolour}

The constituent quark model expressions (\ref{cond2}) and
(\ref{fpi2}) have underpinned much of the folk lore for walking
Technicolour theories\cite{Holdom:1981rm,Appelquist:1986an}. In
brief, in walking Technicolour the gauge coupling is assumed to
transition from perturbative to non-perturbative behaviour at one
scale, $\Lambda_1$ but then the running slows, only crossing some
critical value for inducing chiral symmetry breaking at a scale,
$\Lambda_2$, several orders of magnitude below $\Lambda_1$. In the
period between $\Lambda_1$ and $\Lambda_2$ we imagine that the
anomalous dimension $\epsilon$ of the quark condensate is negative
(so $\bar{q} q$ has dimension less than three) - the condensate
evaluated in the UV is then enhanced taking the rough value
$\Lambda_2^{3-\epsilon}\Lambda_1^{\epsilon}$.

Gap equation analysis\cite{Appelquist:1986an} provides an
alternative but equivalent explanation for the enhancement of the
quark condensate. There walking, which has a larger coupling value
further into the UV, enhances the large $q$ tail of the quark self
energy $\Sigma(q)$. Looking at the constituent quark model
expressions for low energy parameters one can see that $f_\pi$ is
dominated at small $q$ (there is a $q^4$ in the denominator) and
so $f_\pi$ is broadly unchanged by walking. In a Technicolour
model $f_\pi$ sets the W and Z masses and hence the weak scale. On
the other hand the condensate in (\ref{cond2}) is given by a
simple integral over $\Sigma(q)$ and hence grows if the tail of
$\Sigma(q)$ is raised. The condensate is enlarged in walking
theories relative to the weak scale. In extended Technicolour
models\cite{Eichten:1979ah} the condensate determines the standard
model fermion masses - increasing it drives up the extended
Technicolour scale, potentially suppressing flavour physics below
current experimental bounds.

Do our holographic expressions agree with this story? The
challenge is to simulate walking in a holographic setting. The
problem is that we are always at strong coupling (large N) if we
have a weakly coupled gravity dual. As we have seen, the
introduction of any conformal symmetry breaking through the
running coupling causes chiral symmetry breaking\footnote{
Attempts to find backreacted holographic models of gauge theories
with a walking profile such as those in
\cite{Nunez:2008wi,Elander:2009pk} could fall foul of this problem
were they used to generate chiral symmetry breaking.}. We can not
therefore reproduce directly the physics at the scale $\Lambda_1$
discussed above where the theory moved to strong coupling but
without causing chiral symmetry breaking.

As a first attempt to address this point we can be led by the
solutions in Figure 1 as a result of the coupling ansatz in
(\ref{coupling}). If we increase the parameter $\Gamma$ we
effectively smear the scale at which the chiral symmetry breaking
is induced over a range of $r \sim \Gamma^{-1}$. Could we use this
smeared range to represent the separation between $\Lambda_1$ and
$\Lambda_2$ above? The effect of the smearing is to enhance the
tail of the self energy just as expected in walking theories.

If we now turn to the holographic expressions (\ref{fpi}) \&
(\ref{cond}) we see that they naively share the same response to
enhancing the tail of $\Sigma_0$ as the constituent quark model
expressions (\ref{cond2}) \& (\ref{fpi2}) did to raising the tail
of $\Sigma(q)$. In particular again $f_\pi$ has a $1/\rho^4$
factor in the denominator of each integral involved, making it,
one would expect, insensitive to changes in the tail of
$\Sigma_0$. The expression for the condensate though is sensitive
to the tail and should grow as walking is introduced. In fact
though this analysis neglects the dependence of these functions on
the derivatives of the gauge coupling and the self energy function
$\Sigma_0$ - this additional understanding of dynamics coming from
the gauge coupling running lies beyond the constituent quark model
pictures. Both (\ref{fpi}) and (\ref{cond}) are dominated around
the points of maximum change in the coupling and $\Sigma_0$. Note
though that the derivative of the coupling, $\partial_{r^2}
\beta$, is evaluated on the brane, which in the cases above has
precisely embedded itself so as to avoid large derivatives in
$\beta$. By smoothing these functions through increasing $\Gamma$
we include extra functional behaviour. In fact these changes in
the derivatives are more numerically important than the rise in
the tail of $\Sigma_0$ for the plots in Figure 1. This means that
the more ``walking"
 looking self energies in fact give a slightly lower condensate for a fixed
 value of $f_\pi$.  The simple
coupling ansatz in (\ref{coupling}) does
 not therefore accommodate a behaviour we can interpret in the usual
 walking picture. The model
 does suggest that there could be considerable variation in the
 ratio of the condensate to $f_\pi$ in gauge theories with rather
 different speeds of IR running though. A recent lattice analysis
 suggest this ratio could vary as the number of quark flavours is
 changed in QCD \cite{Appelquist:2009ka}.

To take advantage of the similarities between (\ref{fpi}) \&
(\ref{cond}) and (\ref{cond2}) \& (\ref{fpi2}) one would need to
keep the derivatives of the coupling and $\Sigma_0$  roughly fixed as the
scale at which that change occurred was moved out to larger
$\rho$. Our equations would in such a scenario provide the
enhancement of the condensate that one looks for in a walking
theory. Essentially one would want a self energy that rose sharply
at large $\rho$ but then flattened to meet the $w_5$ axis at the
same value as the curves in Figure 1. This in fact matches the
crucial signal of walking that one would expect $\Sigma_0(\rho=0)
\ll \Lambda$ with $\Lambda$ the scale at which the high scale
running occurs. Within holographic models this should be the
crucial signal of walking.

This scenario suggests we are mimicking a slightly different
walking dynamics in the gauge theory than that discussed above -
imagine a theory in which the coupling ran to strong coupling
(call this scale $\Lambda_1$ again) and then entered a conformal
regime with coupling value slightly above the critical value
needed to form a condensate. If the coupling was tuned from above
sufficiently close to the critical value in its conformal window
then a self energy would form but with a size considerably below
$\Lambda_1$.

Realizing this sort of walking behaviour can be done in a
straightforward, if adhoc, fashion. We need to break the symmetry
between $\rho$ and $w_5,w_6$ in the coupling ansatz $\beta$. A
simple ansatz is just to shift our previous ansatz out to larger
$\rho$: \beq \label{walkbeta} \begin{array}{ll} \beta =  a+1 - a
\tanh\left[ \Gamma (\sqrt{(\rho-\lambda_1)^2 + w_5^2 + w_6^2} -
\lambda) \right] &
 \rho \geq \lambda_1 \\
&\\
\beta=1 & \rho < \lambda_1 \end{array}\eeq  This ansatz, which
we sketch in Figure 2, leaves
the derivative of 
$\beta$ unchanged but shifted by
$\lambda_1$ in $\rho$ - this will ensure the condensate, which is 
given by
(\ref{cond}) and dominated around $\lambda_1$ where the derivative
of $\beta$ is non-zero, will grow as $\lambda_1^3$. The embedding will 
still plateau around the same value of $w_5$ since above the step 
(which is quite sharp) the 
space is AdS and the embeddings must be flat.
Below $\lambda_1$ the embedding becomes
flat since the geometry is AdS (the first derivative of $\Sigma_0$
at $\rho=\lambda_1$ is smooth). Obviously this choice of $\beta$ 
below $\lambda_1$ looks
peculiar - one could though imagine that in that region there is a
sharp step function to large coupling at small $w_5,w_6$ - the
embeddings would remain the same.

\begin{center}
$\left. \right.$ \hspace{-3cm}
\includegraphics[width=180mm]{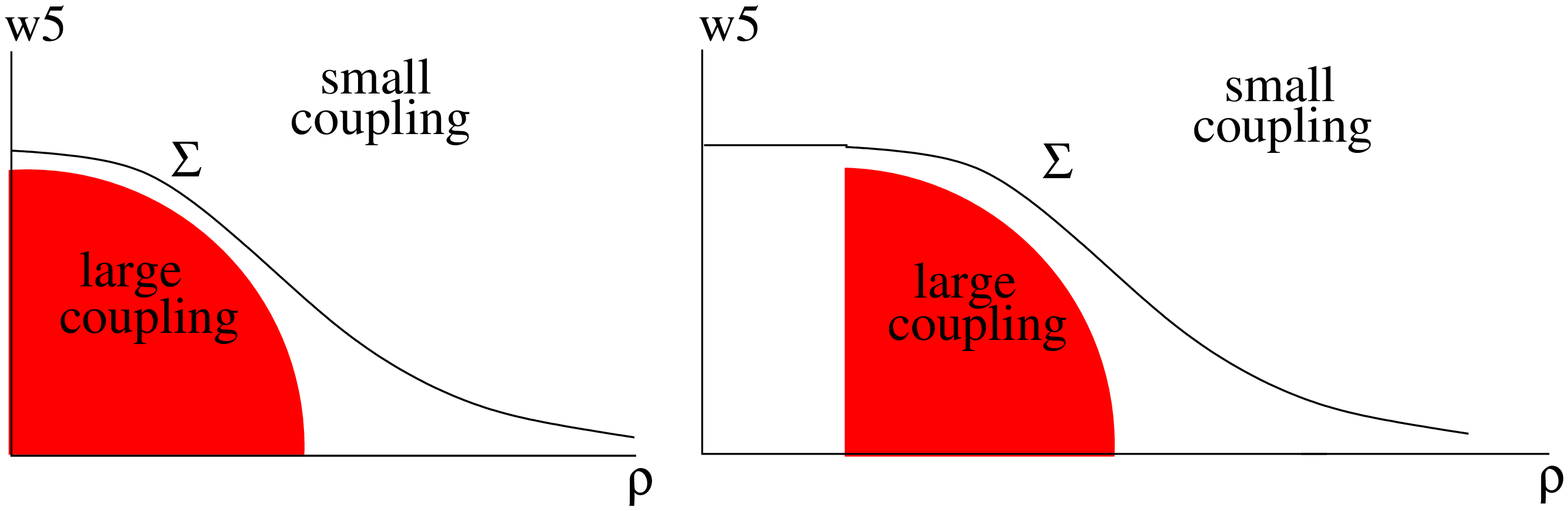}
\end{center} \vspace{-2.5cm}

{\small{Figure 2: A sketch of the area in which the coupling is
large in our ansatz in (\ref{walkbeta}) and the resulting form of
the embeddings $\Sigma_0$ - on the left for $\lambda_1=0$ and on
the right for a non-zero $\lambda_1$ }} \vspace{1cm}

With the embeddings from this walking $\beta$ ansatz we can
analytically see how the expressions for $f_\pi$ and $\langle
\bar{q} q \rangle$ change with $\lambda_1$. In (\ref{fpi}) the
numerator will become independent of $\lambda_1$ as it grows
whilst the denominator, which is proportional to the derivatives
of $\Sigma_0$ and $\beta$ will fall as $1/\lambda_1$. $f_\pi$ will
therefore scale as $\lambda_1^{1/2}$. The condensate expression
(\ref{cond}) is dominated around $\lambda_1$ where the derivative
of $\beta$ is non-zero - it will grow as $\lambda_1^3$. Therefore
if we raise $\lambda_1$ at fixed $f_\pi$ the condensate will grow
as $\lambda_1^{3/2}$. The rise is  consistent with the usual
claims that a walking theory will enhance the condensate.

 It is also possible to numerically confirm this behaviour at least for
small $\lambda_1$. In Figure 3 we show numerical embeddings, displaying
the behaviour shown in Figure 2, as $\lambda_1$ is increased from 0
to 8. To keep the plateau value exactly equal we have tuned $\Gamma$
in the coupling ansatz (it changes from 1 to 3.6 across these plots). 
The condensate grows by an order of magnitude
across these plots and in the large $\lambda_1$ limit will presumably
match the analytic behaviour discussed although more and more tuning
of $\Gamma$ would be needed.  

\begin{center}
\includegraphics[width=70mm]{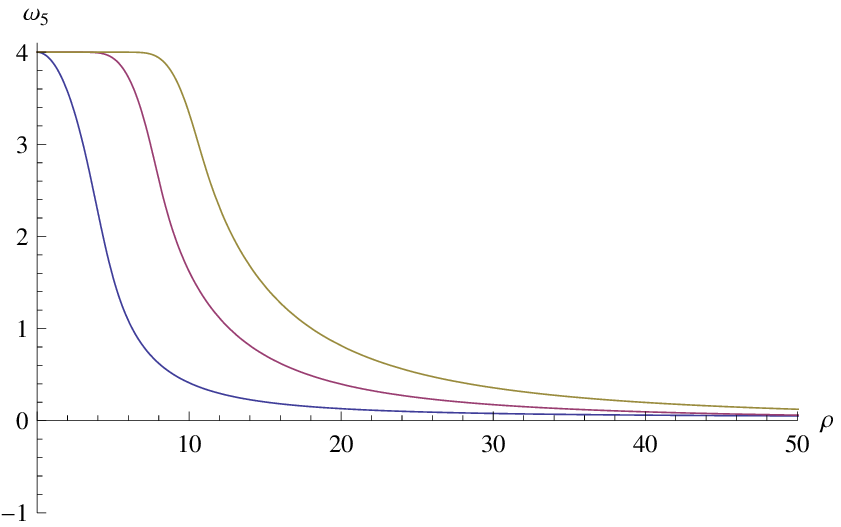}
\end{center} 

{\small{Figure 3: Numerically determined embeddings for the coupling
ansatz in (\ref{walkbeta}). These curves all have $a=3$ and $\lambda=3.19$ 
in addition the curves from from left to right
correspond to the parameter choices $\lambda_1=0, \Gamma=1$,
$\lambda_1=5, \Gamma=3.51$,$\lambda_1=8, \Gamma=3.63$. }} \vspace{1cm}

Note that breaking the symmetry between $\rho$ and $w_5,w_6$ in
the $\beta$ ansatz is still consistent with the symmetries of the
D3-D7 system. In fact interestingly a distinction between the
$\rho$ and $w_5,w_6$ directions is precisely what one would expect
in a geometry backreacted to the D7 branes\cite{Polchinski,
Kirsch:2005uy}. It is therefore plausible that one could fine tune
the number of quark flavours in some D3-D7 system to obtain these
forms of ansatz for the dilaton.

\section{The D3-D5 System}

We now turn to an alternative attempt to describe aspects of
walking dynamics with holography. On first meeting the
D3/(probe)D7 system it seems as if that system should
fundamentally be a walking gauge theory - the ${\cal N}=4$ gauge
dynamics is conformal and strongly coupled in the UV. When we
introduce running in the IR that triggers chiral symmetry
breaking, should the physics not be closer in spirit to that of  a
walking theory rather than QCD? Why did we have to work so hard
above to make that system walk? The reason it is not a walking
theory is that the UV of the D3/D7 system possesses ${\cal N}=2$
supersymmetry which both forbids a quark condensate and protects
the dimension of the $\bar{q}q$ condensate at three. That the self
energy profiles $\Sigma_0$ fall off as $1/\rho^2$ in the analysis
above is driven by that UV supersymmetry and mimics the behaviour
of asymptotically free QCD.

It is natural then to look for a way to introduce quarks into
${\cal N}=4$ super Yang Mills which breaks supersymmetry even in
the far UV. Using a D5 probe to introduce quarks seems the
simplest example to explore. Here we consider the system with a
four dimensional overlap of the D3 and the D5 not a three
dimensional overlap as studied in \cite{Karch:2000gx}. Note that
the strings between the D3 and D5 remain bi-fundamental fields of
the gauge symmetry and global symmetry. The lowest energy modes of
those strings are still at heart the gauge field, that would be
present if the strings were free to move in the whole space, which
become scalar fields, and the gaugino partners that become the
fermionic quarks. In a non-supersymmetric theory the scalars will
most likely become massive leaving fermionic quark multiplets in
the ${\cal N}=4$ theory.

The metric of $AdS_5\times S^5$ can be written in coordinates
appropriate to the D5 embedding as:
\begin{equation}
ds^2=\frac{1}{g_{uv}}
\left[\frac{r^2}{R^2}\eta_{ij}dx^idx^j+\frac{R^2}{r^2}
\left(d\rho^2+\rho^2d\Omega_1^2+d\omega_3^2+d\omega_4^2+d\omega_5^2
+d\omega_6^2\right)\right] \label{eq:}
\end{equation}
with $r^2=\rho^2+\omega_3^2+\omega_4^2+\omega_5^2+\omega_6^2$ and
$\rho^2=\omega_1^2+\omega_2^2$. $R$ is the radius of $AdS$
$R^4=4\pi g_{uv}^2 N \alpha'^2$ The D3 brane is extended in the
$x_i$ dimensions.  The D5-brane will be extended in the $\rho$ and
$\Omega_1$ directions. The $\omega_3,\omega_4,\omega_5$ and
$\omega_6$ are perpendicular to the D5-brane. $g_{uv}^2$ is the
value of the dilaton for $r\rightarrow \infty$.

Let us first analyze the system with a constant dilaton
\begin{equation}
e^{\phi}=g_{uv}^2
    \label{}
\end{equation}
The action for a probe D5 brane assuming the embedding
$\omega_5(\rho), \omega_3=\omega_4=\omega_6=0$ is:
\begin{equation}
\begin{split}
S_{D5}&=-T_5 \int{d^8\xi e^\phi \sqrt{-\det P\left[G\right]_{ab}}}\\
&=-\overline{T_5} \int{d^4x d\rho r^2 \rho \sqrt{1+\left(\partial_\rho w_5\right)^2}},\\
\end{split}
    \label{}
\end{equation}
where $T_5=1/(2\pi)^5 \alpha'^3 $ and $\overline{T_5}=T_5  2\pi
/R^2 g_{uv} $. The embedding equation is
\begin{equation}
\partial_\rho \left[r^2\rho
\frac{\left(\partial_\rho \omega_5 \right)}
{\sqrt{1+\left(\partial_\rho \omega_5\right)^2}}\right]
-2\omega_5 \rho \sqrt{1+\left(\partial_\rho \omega_5\right)^2}=0
    \label{}
\end{equation}
The large $\rho$ behaviour of these solutions is \beq
\label{d5asy} \omega_5 \sim m \rho^{\sqrt{3}-1} +
c/\rho^{1+\sqrt{3}} \eeq The full embeddings are shown on the left
hand of Figure 4. Note that as $m \rightarrow0$ in the UV
asymptotics the full solutions lie along the $\rho$ axis
indicating that the condensate $c=0$ and there is no spontaneous
chiral symmetry breaking - this is a simple result of the absence
of a scale in the conformal field theory.

We continue to interpret the parameter $m$ in the D5 brane
embedding as the quark mass. Then from equation (\ref{d5asy}) we
can see that there is an effective anomalous dimension present for
that mass - its dimension is $2-\sqrt{3}$. The parameter $c$ is
then the quark condensate and has dimension $2 + \sqrt{3}$. the
change in the dimension of these operators in the UV conformal
regime is exactly the physics that underlies the walking idea.
Amusingly though here the anomalous dimension of the quark
condensate is positive rather than negative as usually envisaged
in walking theories. The D3/D5 system will not apparently be much
use for constructing a phenomenological technicolour model. On the
other hand here we are simply interested in testing the intuition
for walking theories so we will continue to investigate for more
formal reasons.

\begin{center}
\includegraphics[width=70mm]{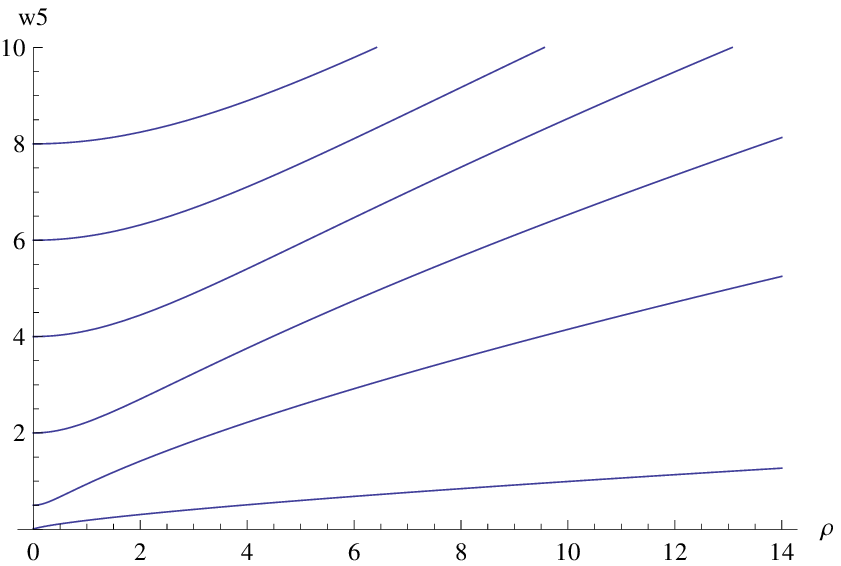} 
\includegraphics[width=70mm]{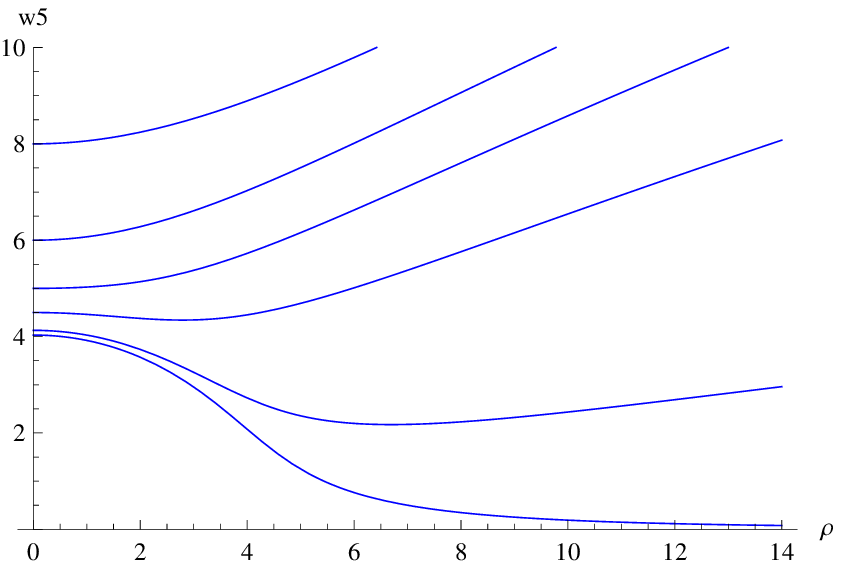}
\end{center}

{\small{Figure 4: The regular embeddings of a D5 brane in pure AdS
with $\beta=1$ on the left. On the right the chiral symmetry
breaking embeddings for the ansatz for $\beta$ in (\ref{coupling})
with $\Gamma=1, \lambda=3, a=5$. }} \vspace{1cm}

\subsection{D3-D5 Embedding with a Non-Trivial Dilaton}

Let us now include a non-trivial dilaton (gauge coupling) profile
as we did above in the D3-D7 system
\begin{equation}
e^\phi=g_{YM}^2(r^2) =g^2_{uv} \beta(r^2).
    \label{}
\end{equation}
For $r\rightarrow \infty$ $\beta \rightarrow 1$. The action is now
\begin{equation}
S_{D5}=-\overline{T_5} \int{d^4x d\rho r^2 \beta \rho
\sqrt{1+\left(\partial_\rho w_5\right)^2}}.
    \label{}
\end{equation}
The embedding equation is
\begin{equation}
\partial_\rho \left[r^2 \beta \rho \frac{\left(\partial_\rho \omega_5 \right)}
{\sqrt{1+\left(\partial_\rho \omega_5\right)^2}}\right]-2\omega_5
\rho \sqrt{1+\left(\partial_\rho \omega_5\right)^2}
\left[\beta+r^2 (\partial_{r^2}\beta)\right]=0
\end{equation}
The embeddings can be seen on the right in Figure 4 for the ansatz
for $\beta$ in (\ref{coupling}). There is again chiral symmetry
breaking with a non-zero $w_5(\rho=0)$ as $m \rightarrow 0$ in the
UV. The self energy curves fall off faster at large $\rho$ which
matches expectations from gap equations in a theory where the
quark condensates dimension grows in the walking regime.

The embedding breaks the SO(4) symmetry in the $\omega_3 -
\omega_6$ directions so we expect there to be Goldstone modes
present. For example, there should be an equivalent solution when
rotating the embedding in e.g. the $\omega_5-\omega_6$ plane.
Let's look at small fluctuations around the embedding $\Sigma_0$
in the $\omega_6$ direction to find a Goldstone boson. The action
for such fluctuations in quadratic order is
\begin{equation}
\begin{split}
S_5=-\overline{T_5}\int d^4x d\rho & r^2 \beta \rho
\sqrt{1+\left(\partial_\rho \Sigma_0\right)^2} \left[1+
(\partial_{r^2}\beta) w_6^2 + \frac{1}{2}\frac{(\partial_\rho
\omega_6)^2}{1+(\partial_\rho \Sigma_0)^2}
+\frac{1}{2}\frac{R^4}{r^4}(\partial_x\omega_6)^2+\ldots\right]\\
    \label{mesonAction}
    \end{split}
\end{equation} where again $r^2, \beta$ and $\partial_{r^2} \beta$
are all evaluated on the the D7 brane world volume $\Sigma_0$.  We
seek fluctuations of the form $\omega_6(x,\rho)=f_n(\rho) e^{i k
\cdot x}$ with $k^2=-M_n^2$. The equation of motion for the
fluctuations give the following equations for $f_n$
\begin{equation}
\partial_\rho \left[r^2 \beta \rho \frac{(\partial_\rho f_n)}
{\sqrt{1+\left(\partial_\rho \Sigma_0\right)^2}} \right]+\frac{R^4}{r^2}
\beta \rho \sqrt{1+\left(\partial_\rho \Sigma_0\right)^2} M_n^2 f_n
- 2 \rho \sqrt{1+\left(\partial_\rho \Sigma_0\right)^2} \left[\beta
+r^2 \partial_{r^2}\beta\right] f_n=0
    \label{GoldEoM}
\end{equation}
The equation with $M^2=0$ and $f_0=\Sigma_0$ is the embedding
equation (\ref{embed}) revealing the presence of the Goldstone
mode.

The Lagrangian for the Goldstone field is found by writing
$\omega_6 = f_0(\rho) \Pi(x) = \Sigma_0 \Pi(x)$ in
(\ref{mesonAction}) and integrating over $\rho$. We can expand
$r^2\beta$ with $r^2=\rho^2+\Sigma_0^2+\Sigma_0^2\Pi^2$ as
$r^2\beta(r^2)=r^2\beta(r^2)|_{r^2=\rho^2+\Sigma_0^2}+
\Sigma_0^2\Pi^2
\partial_{r^2}(r^2\beta)|_{r^2=\rho^2+\Sigma_0^2}$
and then use the equation of motion (\ref{GoldEoM}) to eliminate
the second and third terms in (\ref{mesonAction}) for $M_n=0$.
This procedure gives the Lagrangian to quartic order
\begin{equation}
\begin{split}
\mathcal{L}=&-\overline{T_5}\int   d\rho  r^2 \beta \rho
\sqrt{1+\left(\partial_\rho \Sigma_0\right)^2}\\
& \left. \right.
\left[1+\frac{1}{2}\frac{R^4}{r^4}\Sigma_0^2(\partial_x\Pi)^2
+\frac{1}{4}\frac{R^4}{r^4}\left( \frac{(\partial_\rho \Sigma_0)^2
\Sigma_0^2}{1+(\partial_\rho \Sigma_0)^2}+ 2\Sigma_0^4
\frac{\partial_{r^2}\left( \beta r^2\right) }{\beta r^2}\right)
\left( \partial_x\Pi^2\right)  \Pi^2 +\ldots\right].\\
\label{GoldLagrangian}
\end{split}
\end{equation}

We can now rescale $\Pi$ in (\ref{GoldLagrangian}) and get an
expression for $f_\pi$. We find
\begin{equation}
{f_\pi^2 \over \Lambda^2} ={- 2 N^{1/2} \over \pi^{3/2} \lambda^2}
  \frac{\left[ \int{d\rho \beta \rho \sqrt{1+\left(\partial_\rho
\Sigma_0\right)^2} \frac{\Sigma_0^2}{\rho^2+\Sigma_0^2}} \right]^2
}{\left[ \int
 d\rho{ {\Sigma_0^2 \over (\rho^2 + \Sigma_0^2)^2 } \partial_\rho
 \left( (\rho^2 + \Sigma_0^2)\beta \rho \Sigma_0 (\partial_\rho
 \Sigma_0) \over \sqrt{1+\left(\partial_\rho
\Sigma_0\right)^2} \right)}\right] }
\end{equation}

We also want to find out the value of the quark condensate. We
expand $r^2\beta$ in (\ref{GoldLagrangian}) with $r=\rho^2
+(\Sigma_0+m \rho^{\sqrt{3}-1})^2$ as
$r^2\beta=r^2\beta|_{r^2=\rho^2+\Sigma_0^2} +\partial_{r^2}(r^2
\beta)|_{r^2=\rho^2+\Sigma_0^2}(2 m \rho^{\sqrt{3}-1} \Sigma_0
+{\cal O}(m^2))$. Then we can compare the vacuum energy, $V_0$, in
(\ref{GoldLagrangian}) with the vacuum energy of the chiral
Lagrangian to find the quark condensate
\begin{equation}
 {\langle \overline{q}q \rangle \over \Lambda^{2+\sqrt{3}}}={-N^{1/2}
\over 2 g^2_{uv} N \pi^{1/2} \lambda^{2+\sqrt{3}}} \left. \int
d\rho \rho^{\sqrt{3}} \sqrt{1+\left(\partial_\rho
\Sigma_0\right)^2} \Sigma_0
\partial_{r^2}(r^2 \beta)\right|_{r^2=\rho^2+\Sigma_0^2}.
\end{equation}

These expressions for $f_\pi$ and $\langle \bar{q}q\rangle$ are in
some ways similar to those in the D3/D7 system. $f_\pi$ is again
dominated at low $\rho$ whilst the condensate is more sensitive to
the tail of $\Sigma_0$. In the D5 setting $\Sigma_0$ falls off
more quickly in the UV and will suppress the condensate. This
matches the chiral quark model results. On the other hand the
factor of $N^{1/2}$ before each expression suggests some radical
redistribution of the degrees of freedom in the UV conformal
regime which we can offer no explanation for.

It is important to also note that one can not directly compare the
condensates in the D5 and D7 cases since they have different
intrinsic dimension even in the far UV. In fact to convert the
D3/D5 theory to the usual walking set up would require the
inclusion of extra UV physics (equivalent to that at the scale
$\Lambda_1$ in the walking discussion above) where the
condensate's dimension changes to three. The condensate above that
scale would be suppressed by a further factor of roughly
$\Lambda_1^{\sqrt{3}-1}$.

Whilst the D3/D5 system may not form the basis of any helpful
phenomenological model we do believe that the walking paradigm is
the correct way to interpret the system and the anomalous
dimensions present in the UV.

\section{Conclusions}

We have presented a general description of chiral symmetry
breaking in the D3/D7 system that describes a strongly coupled
gauge theory with quarks. The model allows one to compute the
dependence of the parameters of the low energy chiral Lagrangian
on the running coupling or dilaton form. Our integral formulae for
$f_\pi$ and the quark condensate allow analytic understanding of
how these quantities depend on the coupling and the dynamical mass
of the quark in a similar way to the results of chiral quark
models and the Pagels-Stokar formula. Our model is not complete
since we do not back react the geometry to the dilaton. However,
we view this as a necessary evil to construct intuition in this
type of set up to the response to different dilaton profiles. This
toy environment should provide good guidance for those wishing to
construct fully backreacted solutions that show specific
phenomena.

We have used our results to understand how walking like gauge
dynamics could be included in a holographic framework. The crucial
signal of walking should be that the quark self energy at zero
momentum should be much less than the scale at which conformal
symmetry breaking is introduced. We displayed in figure 2 the form
a dilaton profile must take to achieve walking. Our integral
equations support the usual hypothesis that walking in a gauge
theory would tend to boost the value of the quark condensate
relative to the value of $f_\pi$.

Finally we studied the non-supersymmetric D3/D5 system with a four
dimensional overlap and proposed that the conformal UV of the
theory should be considered as a walking phase of a gauge theory.
The anomalous dimensions of the quark mass and condensate were
computed - in this theory the dimension of the quark condensate is
$2 + \sqrt{3}$ which is greater than the canonical dimension $3$.
Normally walking is constructed to lower this dimension but this
theory hopefully nevertheless adds to our knowledge of walking
behaviour. \vspace{2cm}

\noindent {\bf Acknowledgments:} NE is supported by an STFC
rolling grant. AG and GW are supported by School of Physics and
Astronomy, University of Southampton scholarships. RA thanks
Fundacao ciencia e tecnologia for funding his studentship. We are
grateful for discussions with Ed Threlfall in the early stages of
this work and to Roman Zwicky for comments on the manuscript.

\end{document}


\vspace{2cm}


\begin{thebibliography}{ll}

\bibitem{Polchinski}
  M. Grana and J. Polchinski,
  Phys. Rev. {\bf D65} (2002) 126005,
  [arXiv: hep-th/0106014].

\bibitem{Bertolini:2001qa}
  M.~Bertolini, P.~Di Vecchia, M.~Frau, A.~Lerda and R.~Marotta,
  Nucl.\ Phys.\  B {\bf 621}, 157 (2002)
  [arXiv:hep-th/0107057].

\bibitem{Karch}
  A.~Karch and E.~Katz,
  JHEP {\bf 0206}, 043 (2002)
  [arXiv:hep-th/0205236].



\bibitem{Malda}
J.~M.~Maldacena,  Adv.\ Theor.\ Math.\ Phys.\  {\bf 2}, 231 (1998)
Int.\ J.\ Theor.\ Phys.\  {\bf 38}, 1113 (1999)
[arXiv:hep-th/9711200].

\bibitem{Witten:1998qj}
  E.~Witten,
  Adv.\ Theor.\ Math.\ Phys.\  {\bf 2} (1998) 253
  [arXiv:hep-th/9802150].

\bibitem{Gubser:1998bc}
  S.~S.~Gubser, I.~R.~Klebanov and A.~M.~Polyakov,
  Phys.\ Lett.\  B {\bf 428} (1998) 105
  [arXiv:hep-th/9802109].

\bibitem{Erdmenger:2007cm}
  J.~Erdmenger, N.~Evans, I.~Kirsch and E.~Threlfall,
  Eur.\ Phys.\ J.\  A {\bf 35} (2008) 81
  [arXiv:0711.4467 [hep-th]].




\bibitem{Maldacena:1998im}
  J.~M.~Maldacena,
  Phys.\ Rev.\ Lett.\  {\bf 80} (1998) 4859
  [arXiv:hep-th/9803002].

\bibitem{Rey:1998ik}
  S.~J.~Rey and J.~T.~Yee,
  Eur.\ Phys.\ J.\  C {\bf 22} (2001) 379
  [arXiv:hep-th/9803001].

\bibitem{Mateos}
 M.~Kruczenski, D.~Mateos, R.~C.~Myers and D.~J.~Winters,
 JHEP {\bf 0307} 049, 2003
 [arXiv:hep-th/0304032].

\bibitem{Son:2007vk}
  D.~T.~Son and A.~O.~Starinets,
  Ann.\ Rev.\ Nucl.\ Part.\ Sci.\  {\bf 57} (2007) 95
  [arXiv:0704.0240 [hep-th]].

\bibitem{Gubser:2009md}
  S.~S.~Gubser and A.~Karch,
  arXiv:0901.0935 [hep-th].

\bibitem{Mateos:2007vn}
  D.~Mateos, R.~C.~Myers and R.~M.~Thomson,
  JHEP {\bf 0705} (2007) 067
  [arXiv:hep-th/0701132].




\bibitem{Babington}
  J.~Babington, J.~Erdmenger, N.~J.~Evans, Z.~Guralnik and I.~Kirsch,
  Phys.\ Rev.\  D {\bf 69} (2004) 066007
  [arXiv:hep-th/0306018].

\bibitem{Ghoroku:2004sp}
  K.~Ghoroku and M.~Yahiro,
  Phys.\ Lett.\  B {\bf 604} (2004) 235
  [arXiv:hep-th/0408040].

\bibitem{Kruczenski:2003uq}
  M.~Kruczenski, D.~Mateos, R.~C.~Myers and D.~J.~Winters,
  JHEP {\bf 0405} (2004) 041
  [arXiv:hep-th/0311270].

\bibitem{Filev:2007gb}
  V.~G.~Filev, C.~V.~Johnson, R.~C.~Rashkov and K.~S.~Viswanathan,
  JHEP {\bf 0710} (2007) 019
  [arXiv:hep-th/0701001].

\bibitem{Pagels:1979hd}
  H.~Pagels and S.~Stokar,
  Phys.\ Rev.\  D {\bf 20} (1979) 2947.

\bibitem{Holdom:1990iq}
  B.~Holdom, J.~Terning and K.~Verbeek,
  Phys.\ Lett.\  B {\bf 245} (1990) 612.

\bibitem{Holdom:1981rm}
  B.~Holdom,
  Phys.\ Rev.\  D {\bf 24}, 1441 (1981).

\bibitem{Appelquist:1986an}
  T.~W.~Appelquist, D.~Karabali and L.~C.~R.~Wijewardhana,
  Phys.\ Rev.\ Lett.\  {\bf 57} (1986) 957.

\bibitem{Weinberg:1975gm}
  S.~Weinberg,
  Phys.\ Rev.\  D {\bf 13} (1976) 974.

\bibitem{Susskind:1978ms}
  L.~Susskind,
  Phys.\ Rev.\  D {\bf 20} (1979) 2619.

\bibitem{Sakai:2004cn}
  T.~Sakai and S.~Sugimoto,
  Prog.\ Theor.\ Phys.\  {\bf 113} (2005) 843
  [arXiv:hep-th/0412141].

\bibitem{Coleman:1969sm}
  S.~R.~Coleman, J.~Wess and B.~Zumino,
  Phys.\ Rev.\  {\bf 177} (1969) 2239.


\bibitem{Eichten:1979ah}
  E.~Eichten and K.~D.~Lane,
  Phys.\ Lett.\  B {\bf 90} (1980) 125.



\bibitem{Nunez:2008wi}
  C.~Nunez, I.~Papadimitriou and M.~Piai,
  arXiv:0812.3655 [hep-th].

\bibitem{Elander:2009pk}
  D.~Elander, C.~Nunez and M.~Piai,
  arXiv:0908.2808 [hep-th].


\bibitem{Appelquist:2009ka}
  T.~Appelquist {\it et al.},
  arXiv:0910.2224 [Unknown].

\bibitem{Kirsch:2005uy}
  I.~Kirsch and D.~Vaman,
  Phys.\ Rev.\  D {\bf 72} (2005) 026007
  [arXiv:hep-th/0505164].


\bibitem{Karch:2000gx}
  A.~Karch and L.~Randall,
  JHEP {\bf 0106} (2001) 063
  [arXiv:hep-th/0105132].







\end{thebibliography}
\end{document}